
\vsize=7in
\hsize=5in
\centerline{DIFFERENTIAL \enskip SUBSTITUTIONS \enskip AND}
\centerline{SYMMETRIES \enskip OF \enskip HYPERBOLIC \enskip EQUATIONS}
\bigskip
\centerline{\tenrm S. Ya. Startsev}
\centerline{\sevenrm Mathematical Institute of Ufa Center of Russian Academy,}
\centerline{\sevenrm Chernyshevsky Street 112, 450000 Ufa, Russia}
\centerline{\sevenrm e-mail:  starts\@nkc.bashkiria.su}
\bigskip
{\sevenbf Abstract:} {\sevenrm There are considered  differential
substitutions of the form} $v=P(x,u,u_{x})$ {\sevenrm for which there exists
a  differential operator} $H = \sum^{k}_{i=0} \alpha _{i} D^{i}_{x}$
{\sevenrm such  that the differential substitution maps  the equation}
$u_{t}=H[s(x,P,D_{x}(P),...,D^{m}_{x}(P))]$ {\sevenrm into  an  evolution
equation  for  any  function} $s$ {\sevenrm and any nonnegative integer}
$m${\sevenrm. All  differential  substitutions of the form} $v=P(x,u,u_{x})$
{\sevenrm known  to  the author {\sevenrm have this property. For example,
the  well-known  Miura transformation} $v = u_{x}-u^{2}$ {\sevenrm maps any
equation of the form} \par \centerline{$u_{t}=(D^{2}_{x}+2uD_{x}+2 u_{x})
[s(x,u_{x}-u^{2},D_{x}(u_{x}-u^{2} ),...,D^{k}_{x}(u_{x}-u^{2}))]$}
\par
\noindent {\sevenrm into the equation}
\par
\centerline{$v_{t}=(D^{3}_{x}+4vD_{x}+2v_{x})[s(x,v,{{\partial
v}\over{\partial x}},...,{{\partial^{k} v}\over{\partial x^{k}}})].$} \par
\noindent {\sevenrm The complete classification of such differential
substitutions is  given.  An infinite  set  of  the  pairwise nonequivalent
differential substitutions with the property mentioned above is constructed.
Moreover, a general result about  symmetries  and invariant functions of
hyperbolic  equations is obtained (see theorem 1).}
\par
\bigskip
{\tenrm Let us consider a differential  substitution  of  the  form}
$$v=P(x,u,u_{x}) \eqno{(1)}$$
{\tenrm which maps solutions of an evolution equation into solutions of
another evolution equation. It is sufficient for our aim to give  the
following definition of the differential substitution ( see [1] for  more
general definition ).}
\par {\bf Definition 1:} {\it (1) is called a
differential substitution from an equation} $$u_{t}=f(x,u,u_{1},...,u_{n})
\eqno{(2)}$$ {\it into an equation}
$$v_{t}=g(x,v,v_{1},...,v_{n}) \eqno{(3)}$$
{\it if } $P_{u_{x}}\not=0$ {\it and the relationship}
$$(P_{u_{x}}D_{x}+P_{u})(f)=g(x,P,D_{x}(P),...,D^{n}_{x}(P))$$
\par \noindent {\it is satisfied. Here} $u_{i}$ {\it and} $v_{i}$ {\it are
the} {\it i-th  partial derivatives of\ } $u$ {\it and} $v$ {\it with
respect to} $x$ {\it , respectively,}  $D_{x}$ {\it is the  total
derivative with respect to} $x$ {\it expressed in the variables} $x,u, u_{i}$
{\it as}
$$D_{x}={\partial\over{\partial x}}+u_{1}{\partial\over{\partial
u}}+\sum^{\infty}_{i=1} u_{i+1}{\partial\over{\partial u_{i}}}\enskip .$$
\par
{\tenrm It should be
noted that differential  substitutions  are determined  up   to invertible
(contact and   point) transformations of the equations (2) and  (3).  A
differential substitution} $\pi _{1}$ {\tenrm is said  to  be} {\it
equivalent}  {\tenrm to  a  differential substitution} $\pi _{2}$ {\tenrm if}
$\pi _{1}$ {\tenrm =} $\rho _{2}\circ \pi _{2}\circ \rho _{1}${\tenrm ,
where} $\rho _{1}$ {\tenrm and} $\rho _{2}$ {\tenrm are invertible
transformations of the equations (2) and (3), respectively.  It is shown in
[1] that any differential substitution}
$$\eqalign{v&=\psi(x,u,u_{1},u_{2}) \cr
\overline{x}&=\varphi(x,u,u_{1},u_{2}) \cr}$$
{\tenrm which satisfies the conditions}
$$\varphi_{u_{2}}D_{x}(\psi)=\psi_{u_{2}}D_{x}(\varphi),\qquad
{{\partial^{2}}\over{\partial u^{2}_{2}}}
\left({{\varphi_{u}D_{x}(\psi)-\psi_{u}D_{x}(\varphi)}\over
{\varphi_{u_{1}}D_{x}(\psi)-\psi_{u_{1}}D_{x}(\varphi)}}\right)=0 \eqno{(4)}$$
{\tenrm is equivalent to a differential substitution of  the  form  (1).
Therefore, our study of the differential substitutions (1) is the
study of the more wide class of differential substitutions.  For
example, all substitutions of the form}
$$\eqalign{v&=\psi (x,u,u_{1}) \cr
\overline{x}&=\varphi (x,u,u_{1}) \cr}$$
{\tenrm satisfy the conditions (4).}
\par
{\tenrm We shall assume in this paper  that  for  the  differential
substitution (1) there exists a differential operator}
$$H = \sum^{k}_{i=0} \alpha _{i} D^{i}_{x}$$
{\tenrm where} $k \ge 0, \alpha _{i}$ {\tenrm are functions of} $x,u,u_{i}$
{\tenrm and} $\alpha _{k} \not=0 $ {\tenrm , such  that the differential
substitution transforms the equation}
$$u_{t}=H[s(x,P,D_{x}(P),...,D^{m}_{x}(P))]$$
{\tenrm into  an  evolution equation  for  any  function} $s$   {\tenrm
and any nonnegative integer} $m$ {\tenrm . All differential substitutions
(1) known to the author have this property.}
\par
{\bf Examples:} {\tenrm The example of the Miura transformation is contained
in Abstract. In this case } $H=D^{2}_{x}+2uD_{x}+u_{x}.$
\par
{\tenrm Other examples are:}
$$\xalignat 2
&v=u_{1},& &H=1;\\
&v=u_{1}+e^{u},& &H=D_{x}+u_{1};\\
&v=u_{1}+e^{u}+e^{-u},& &
H=D^{2}_{x}+{{u_{1}+e^{-2u}-e^{2u}}\over{u_{1}+e^{u}+e^{-u}}} D_{x}+ \\
&&&+{{u_{3}-2u_{1}(e^{2u}+e^{-2u})}\over{u_{1}+e^{u}+e^{-u}}}-
{\left({{u_{2}+u_{1}(e^{u}-e^{-u})}\over{u_{1}+e^{u}+e^{-u}}}\right)}^{2}-
u_{1}^{2} \enskip . \\
\endxalignat
$$
{\tenrm It should be noted that  these  examples
show very well  that integrability of an equation and  existence  of  a
differential substitution relating this equation to another equation are  not
the same.}
\par
{\tenrm The present work is based on  the  fact  that  (1) is
the differential  substitution from  the  equation  (2)   into   an evolution
equation if and only if the right-hand  side  of  the equation (2) is a
symmetry of the hyperbolic equation}
$$u_{xy}={{-P_{u}}/{P_{u_{1}}}} u_{y}.\eqno(5)$$
{\tenrm This fact was proved in [1].}
\par
{\tenrm We remind that a function} $f$ {\tenrm depending on} $x,u,u_{i},
w_{i}={{\partial^{i} u}\over{\partial y^{i}}}$
{\tenrm is called} {\it a symmetry} {\tenrm of the equation}
$$u_{xy}=F(x,y,u,u_{x},u_{y}) \eqno{(6)}$$
{\tenrm if} $f$ {\tenrm satisfies the condition}
$$\left({D_{x} D_{y}-F_{u_{x}}D_{x}-F_{u_{y}}D_{y}-F_{u}}\right)(f)=0,$$
{\tenrm where} $D_{x}$ {\tenrm and} $D_{y}$ {\tenrm are the total derivatives
with respect to} $x$ {\tenrm and} $y$ {\tenrm by virtue of the
equation  (6), respectively.  It should  be noted that we eliminate the
derivatives} ${\partial^{i+j} u}\over{\partial x^{i} \partial y^{j}}$
{\tenrm with } $i \cdot j \not=0$ {\tenrm by virtue of the equation (6) and
its differential consequences.  Therefore, we assume from now on that all
functions depend on the variables} $x,y,u,u_{i},w_{i}${\tenrm . The total
derivatives} $D_{x}$ {\tenrm and}  $D_{y}$ {\tenrm are expressed in these
variables as}
$$D_{x}={\partial\over{\partial x}}+u_{1}{\partial\over{\partial u}}+
\sum^{\infty }_{i=1}\left(u_{i+1}{\partial\over{\partial u_{i}}}
+D^{i-1}_{y}(F){\partial\over{\partial w_{i}}}\right)$$
$$D_{y}={\partial\over{\partial y}}+w_{1}{\partial\over{\partial u}}+
\sum^{\infty }_{i=1}\left(w_{i+1}{\partial\over{\partial w_{i}}}
+D^{i-1}_{x}(F){\partial\over{\partial u_{i}}}\right)$$
\par
{\bf Definition 2:} {\it A function } $Q$ {\it is  said to be }
$D_{x}${\it-invariant ( } $D_{y}${\it-invariant ) if } $D_{x}(Q)=0$
{\it ( } $D_{y}(Q)=0$ {\it ).  If } $Q$ {\it depend on only } $y$ {\it (
only } $x$ {\it ), then }  $Q$  {\it is called  a  trivial }
$D_{x}${\it-invariant ( } $D_{y}${\it-invariant ) function.}
\par
{\tenrm It is important to observe that any function of } $x,P,D^{i}_{x}(P)$
{\tenrm is } $D_{y}${\tenrm-invariant for the equation (5) and the converse
statement also is true. Using this fact, our assumption and the following
theorem, we can  reduce  the classification problem  for  the differential
substitutions (1) to the classification problem for the equations  (5)  which
admit  a nontrivial } $D_{x}${\tenrm-invariant function.}
\par
{\bf Theorem 1:} {\it Let an equation  (6)  admits  a nontrivial }
$D_{y}${\it-invariant function and there exist a differential operator}
$$H=\sum^{k}_{i=0}\alpha_{i}D^{i}_{x}$$
{\it such that} $H(s)$ {\it is the symmetry of  the  equation
(6)  for any } $D_{y}${\it-invariant  function } $s.$ {\it Then
the equation  (6) admits   a nontrivial } $D_{x}${\it-invariant
function and there exist a differential operator}
$$R=\sum^{m}_{i=0} \beta _{i} D^{i}_{y}$$
{\it such that} $R(q)$ {\it is the symmetry of  the  equation  (6)  for
any } $D_{x}${\it-invariant function} $q.$
\par
{\it Here } $\alpha_{i},\beta_{i}$ {\it are functions of } $x,y,u$ {\it and
the derivatives of } $u$ {\it; } $k,m\ge0,\alpha_{k},\beta_{m}\not=0.$
\par
{\tenrm The main idea of the proof
is to demonstrate finiteness  of the Laplace } {\it Y}{\tenrm-index of the
equation (6) ( the definition of  the Laplace indices is  contained in [2] ).
The latter implies existence of the } $D_{x}${\tenrm-invariant function and
the operator } $R$ {\tenrm ( see [3]-[5] ).  The technique of the  proof  of
finiteness  of the Laplace } {\it Y}{\tenrm-index is similar to one which
is used in [2].}
\par
{\tenrm Let } $f$ {\tenrm be a symmetry of the equation
(6).  It  is easy to verify that the vector field}
$$\partial_{f}=f{\partial\over{\partial u}}+\sum^{\infty }_{i=1} \left(
D^{i}_{y}(f){\partial\over{\partial w_{i}}}+ D^{i}_{x}(f)
{\partial\over{\partial u_{i}}}\right)$$
{\tenrm commutes with the  total derivatives} $D_{x}$ {\tenrm and }
$D_{y}${\tenrm .  Hence, the operator}
$$Q_{*}={{\partial Q}\over{\partial u}}+\sum^{\infty}_{i=1}\left(
{{\partial Q}\over{\partial w_{i}}}D^{i}_{y}+
{{\partial Q}\over{\partial u_{i}}}D^{i}_{x}\right)$$
{\tenrm maps any symmetry into a } $D_{x}${\tenrm-invariant ( }
$D_{y}${\tenrm-invariant ) function if } $Q$ {\tenrm is}
$D_{x}${\tenrm-invariant ( } $D_{y}${\tenrm-invariant ).
Let  us consider  the equation (5).  The straightforward calculation shows
that } $w_{1}g$  {\tenrm is the symmetry of (5) for any}
$D_{x}${\tenrm-invariant functions} $g${\tenrm .  Therefore, if} $Q$ {\tenrm
is} $D_{x}${\tenrm-invariant, then the operator }
$Q_{*}\circ w_{1}=\sum^{k}_{i=0}c_{i}D^{i}_{y}$ {\tenrm maps any}
$D_{x}${\tenrm-invariant function again into a } $D_{x}${\tenrm-invariant
function and the coefficient }  $c_{i}$ {\tenrm must  be}
$D_{x}${\tenrm-invariant.  Taking  this  into account we can easily prove
the  following result  which  was obtained in a more complicated way in [6].}
\par
{\bf Proposition 1:} {\it Let an equation}
$$u_{xy}=b(x,u,u_{x}) u_{y}$$
{\it has a nontrivial} $D_{x}${\it-invariant function. Then  this
equation can be reduced to another  of  the  same  form  admitting  the}
$D_{x}${\it-invariant function}
$$Q={{w_{3}}\over{w_{1}}}-{3\over 2} {\left({{w_{2}}\over{w_{1}}}\right)}^{2}
\eqno(7)$$
{\it by a transformation}
$$u=\psi (x,u). \eqno(8)$$
\par
{\tenrm The equation (5)
has a nontrivial} $D_{x}${\tenrm-invariant function  by our assumption
and theorem 1.  Conversely, if  the  equation  (5) admits  a nontrivial}
$D_{x}${\tenrm-invariant function, then there  exist  a differential
operator} $H=\sum^{k}_{i=0} \alpha _{i} D^{i}_{x}$ {\tenrm
which maps any}  $D_{y}${\tenrm-invariant function of the equation (5)
into the  right-hand  side  of  an evolution equation admitting the
substitution (1) ( see  [4]  ).  Transformations (8) correspond to point
transformations  of  the equation  (2).   Thus differential substitutions (1)
interesting for us are equivalent to ones which correspond to the
equation (5) admitting the} $D_{x}${\tenrm-invariant function (7).}
\par
{\tenrm It is convenient for further consideration  to  express}  $u_{x}$
{\tenrm from (1) and rewrite the  differential  substitution  (1)  as  a
system}
$$\biggl\{ \eqalign{u_{x}&=a(x,u,v) \cr
            v_{y}&=0 \enskip . \cr}\eqno(9)$$
{\tenrm We eliminate all partial derivatives of } $u$ {\tenrm
with respect to } $x$ {\tenrm by virtue of (9). The total
derivatives } $D_{x}$ {\tenrm and } $D_{y}$ {\tenrm are expressed  in
the remaining variables} $x,u,v,v_{i},w_{i}$ {\tenrm as}
$$\eqalign {D_{x}&={\partial\over{\partial x}}+a{\partial\over{\partial u}}+
v_{1} {\partial\over{\partial v}}+\sum^{\infty }_{i=1}\left(v_{i+1}
{\partial\over{\partial v_{i}}}+D^{i}_{y}(a)
{\partial\over{\partial w_{i}}}\right) \cr
D_{y}&={\partial\over{\partial y}}+w_{1}{\partial\over{\partial u}}+
\sum^{\infty }_{i=1} w_{i+1}{\partial\over{\partial w_{i}}}\enskip.\cr}$$
{\tenrm It is easy to see that} $D_{x}\left(
{{w_{3}}\over{w_{1}}}-{3\over 2} {\left({{w_{2}}\over{w_{1}}}\right)}^{2}
\right) = 0$
{\tenrm if and only if} $a_{uuu}=0.$
\par
{\tenrm Thus we obtain the next}
\par
{\bf Theorem 2:} {\it The differential substitution (1) satisfies  our
assumption if and only if it is equivalent to a substitution  of
the same form corresponding to the system (9) with}
$a=\alpha(x,v) u^{2}+\beta(x,v)u+\gamma(x,v).$
\par
\bigskip
\bigskip
{\tenrm To understand how many essentially different  substitutions
are described by the system (9) with } $a_{uuu}=0$ {\tenrm , we need study
the equivalence problem for the substitutions (1). It is  convenient to do in
terms of the system (9).} \par {\tenrm Transformations}
$$\eqalign{x&=\alpha(\overline{x}) \cr
u&=\varphi (\overline{x},\overline{u}) \cr
v&=\psi (\overline{x},\overline{v}) \cr} \eqno(10)$$
{\tenrm of the system  (9)  preserve  the  form  of  the  latter.  These
transformations correspond to the point transformations  of  the
equations  (2)  and  (3)  which  preserve  the   form   of   the
substitution (1). Contact transformations of the  equations  (2)
and (3) preserving the form of (1) induce  a  transformation  of
the system (9)}
$$\eqalign{x&=\varphi (\overline{x},\overline{v}),\varphi_{v} \not= 0 \cr
u&=\psi (\overline{x},\overline{u},\overline{v}) \cr
v&=\Phi (\overline{x},\overline{v}), \cr} \eqno(11)$$
{\tenrm where} $\varphi,\psi$ {\tenrm and} $\Phi $ {\tenrm satisfy the
condition}
$$\psi_{v}-a(\varphi,\psi,\Phi)\varphi_{v}=0\eqno(12)$$
{\tenrm The transformation (11)-(12) preserves the form of (9) and the
right-hand side } $\overline{a}$ {\tenrm  of the transformed system is
expressed  by the formula}
$$\overline{a}={{\varphi_{x}\psi_{v}-\varphi_{v}\psi_{x}}\over
{\varphi_{v}\psi_{u}}}\enskip .\eqno(13)$$
\par
{\tenrm Using the formulas (12) and (13)  we  can  prove  the
next proposition by a simple but tedious argument.}
\par
{\bf Proposition 2:}{\it Let a  transformation  (10)-(13)  relates  a system}
$$\biggl\{ \eqalign{u_{x}&=a(x,u,v)=
\alpha(x,v) u^{2}+\beta(x,v)u+\gamma(x,v) \cr
v_{y}&=0 \cr},$$
{\it where}
${\left(a_{uv}/a_{v}\right)}_{v} \not=0$ {\it , to a system} $$\biggl\{
\eqalign{\overline{u}_{\overline{x}}&=\overline{a}(\overline{x},\overline{u},
\overline{v})=\overline{\alpha}(\overline{x},\overline{v})\overline{u}^{2}+
\overline{\beta}(\overline{x},\overline{v})\overline{u}+\overline{\gamma}(
\overline{x},\overline{v}) \cr
\overline{v}_{y}&=0 \cr},$$
{\it where}
${\left(\overline{a}_{\overline{u}\overline{v}}/\overline{a}_{\overline{v}}
\right)}_{\overline{v}}\not=0.$
{\it Then this transformation must be either  of the form}
$$\eqalign {x&=\varphi(\overline{x},\overline{v}) \cr
v&=\Phi(\overline{x},\overline{v}) \cr
u&={{\lambda(\overline{x},\overline{v})}\over{\overline{u}+\eta(\overline{x},
\overline{v})}}+\xi(\overline{x},\overline{v})\enskip ,\cr} \eqno(14)$$
{\it where} $\lambda,\eta$ {\it and} $\xi$ {\it satisfy the conditions}
$$\eqalign{\eta_{\overline{v}}&=-\lambda\varphi_{\overline{v}}\alpha(\varphi,
\Phi) \cr
\eta_{\overline{x}}&=\eta \overline{\beta}-\eta^{2} \overline{\alpha} -
\overline{\gamma}-\lambda\varphi_{\overline{x}}\alpha(\varphi,\Phi) \cr
\lambda_{\overline{v}}&=\lambda\varphi_{\overline{v}}\left(\beta(\varphi,\Phi)
+2 \xi \alpha(\varphi,\Phi)\right) \cr
\lambda_{\overline{x}}&= \lambda\left(\varphi_{\overline{x}}
(2\alpha(\varphi,\Phi)\xi+\beta(\varphi,\Phi))
+ \overline{\beta} - 2\eta \overline{\alpha} \right) \cr
\xi_{\overline{v}}&=\varphi_{\overline{v}}\left(\alpha(\varphi,\Phi) \xi^{2}
+\beta(\varphi,\Phi)\xi + \gamma(\varphi,\Phi)\right) \cr
\xi_{\overline{x}}&=\varphi_{\overline{x}}\left(\alpha(\varphi,\Phi)\xi^{2}+
\beta(\varphi,\Phi)\xi +\gamma(\varphi,\Phi)\right) +
\lambda \overline{\alpha}, \cr }$$
{\it or of the form}
$$\eqalign{x&=\varphi(\overline{x}) \cr
v&=\Phi(\overline{x},\overline{v}) \cr
u&=\lambda(\overline{x})\overline{u}+\xi(\overline{x}).\cr}\eqno(15)$$
{\it In both cases} $\lambda \not=0.$
\par
{\tenrm We need the above proposition to prove the following} \par {\bf
Proposition 3:} {\it The differential substitutions described  by the system
(9) with} $a=u^{2}+ u v + v^{\gamma},$ {\it where} $\gamma $ {\it is
constant, } $\gamma \not = 0, 1,$ {\it  are pairwise nonequivalent for
different} $\gamma.$ \par {\tenrm To prove this proposition it is sufficient
to verify that the system}
$$\eqalign{\eta_{\overline{v}}&=-\lambda\varphi _{\overline{v}} \cr
\eta_{\overline{x}}&=\eta \overline{v} -\eta^{2} -{\overline{v}}^{r} -\lambda
\varphi_{\overline{x}} \cr
\lambda_{\overline{v}}&=\lambda\varphi_{\overline{v}} (\Phi+2\xi) \cr
\lambda_{\overline{x}}&=\lambda
\left(\varphi_{\overline{x}}(\Phi+2\xi)+\overline{v}-2\eta \right) \cr
\xi_{\overline{v}}&=\varphi_{\overline{v}}(\xi^{2}+\Phi\xi+\Phi^{s}) \cr
\xi_{\overline{x}}&=\varphi_{\overline{x}}(\xi^{2}+\Phi\xi+\Phi^{s})+\lambda
\cr}$$
{\tenrm is incompatible if} $r \not=s.$ {\tenrm Then it should  be
checked that the differential substitutions of the above form  are
nonequivalent under point transformations (15).}
\par {\tenrm In conclusion,
we give without a proof the next  results which are obtained by the more
detailed study of the equivalence problem for the substitutions (1).} \par
{\bf Proposition 4:} {\it The  differential  substitution   (1) is equivalent
to  a differential substitutions   of   the   form} $v=u_{x} + p(x,u)$ {\it
if and only if the right-hand  side}  $a$ {\it of  the corresponding  system
(9) satisfy   one of   the   following conditions:}
$${\left({{a_{vv}}\over{a_{v}}}\right)}_{u}=0 \qquad
or \qquad
{\left({{{\left({{a_{xv}+ a a_{uv}}\over{a_{v}}}-a_{u}\right)}_{u}}\over
{{\left({{a_{vv}}\over{a_{v}}}\right)}_{u}}}\right)}_{u}=0$$
\par
{\bf Theorem 3:} {\it Any differential substitution (1)  described  by the
system (9) with} $a_{uuu}=0$ {\it is  equivalent  to one  of  the following
substitutions:}
\par
{\it A) } $v=u_{x}$
\par
{\it B) } $v=u_{x}+e^{u}$
\par
{\it C) } $v=u_{x}+e^{u}+e^{-u}$
\par
{\it D) } $v=u_{x}+u^{2}$
\par
{\it F) the substitutions  described  by  the system (9)  with}
$a=u^{2} + v u + \gamma(x,v),\gamma_{vv}\not=0.$
\par
{\it The cases A,B,C,D and F are nonequivalent.}
\par
{\tenrm Using proposition 4 we easily verify that the case F can not be
reduced to the form } $v=u_{x}+p(x,u).$ {\tenrm Thus there exist four
differential substitutions of the form} $v=u_{x}+p(x,u)$
{\tenrm and the
infinite set of substitutions which are not equivalent to any substitution }
$v=u_{x}+p(x,u).$
\par
\medskip
\centerline{\bf Acknowledgments} \par {\tenrm The author wishes to thank  S.
I. Svinolupov for many  useful discussions.  The author  especially thanks V.
V.  Sokolov who offered the idea of this research  and recommended to apply
the technique of  the generalized  Laplace transformations to the present
work.  The idea of the proof of proposition 1 also belongs to V. V. Sokolov.}
\par \medskip \centerline{\bf References.} \par {\tenrm [1]  V.  V.
Sokolov.}  {\it On the symmetries of evolutions equations.}{\tenrm  Russian
Math. Surv., 43(5), 1988, 165-204.} \par {\tenrm [2] I. M. Anderson and N.
Kamran.} {\it The variational bicomplex for second order scalar partial
differential  equations in  the plane.}{\tenrm CMR technical report,
September, 1994.} \par {\tenrm [3] V. V. Sokolov and A. V. Zhiber.} {\it On
the Darboux  integrable hyperbolic equations.}{\tenrm  To appear in Phis.
Lett. A.} \par {\tenrm [4] A. V. Zhiber, V. V. Sokolov and S.  Ya.
Startsev.}  {\it On the Darboux integrable nonlinear hyperbolic
equations.}{\tenrm  To appear in Doklady RAN ( in Russian ).} \par {\tenrm
[5]  M.  Juras  and  I.M  Anderson.}  {\it Generalized  Laplace invariant and
the Method of  Darboux.} {\tenrm Preprint,  University  of Montreal, June,
1995.} \par {\tenrm [6] A. V. Zhiber.}  {\it Quasilinear  hyperbolic
equations  with an infinite-dimensional algebra of symmetries.} {\tenrm
Doclady RAN, v.58, No 4, 1994 ( in Russian ).}
\par
\end{document}